# An electrically tunable metaatom for visible light

Janna Wilhelmsen[1], Longzhu Liu[1], Harry Miyosi Silalahi[1], Suraya Kazi[1], Giancarlo Cincotti[1], Shangzhi Chen[1], Dongqing Lin[1], Yulong Duan[1], Magnus P. Jonsson[1,*]

[1] Laboratory of Organic Electronics, Department of Science and Technology, Linköping University; Norrköping, SE-60174, Sweden

*Corresponding author. Email: magnus.jonsson@liu.se

**Phase-gradient metasurfaces provide powerful wavefront control through two-dimensional arrangement of nanostructures acting as metaatoms[1–3]. While dynamic tuning forms a major driver for future breakthroughs and applications in this area[4], current metaatoms are generally static or limited to operation in the infrared[5,6]. Here, we present a metaatom that is both electrically tunable and operates in the visible. Its function originates from an excitonic absorption band of a dedoped conducting polymer, which together with low background permittivity induces optical metallicity in a selected part of the visible. This allows anisotropic nanostructures to support excitonic resonances along one direction and not the other, promoting polarization-dependent optical response which can be toggled off and on through reversible doping induced by small bias potentials. Our study details the mechanism of these metaatoms and demonstrate their use in electrically tunable phase-gradient metasurfaces for visible light, including erasable and rewritable holograms.**

## Introduction

Phase-gradient metasurfaces have transformed the control of optical wavefronts, with applications ranging from flat ultrathin lenses[1] and holograms[3] to advanced polarization control[2]. A major frontier in this rapidly advancing field is dynamic tunability, which enable active control over optical functionalities and expand the application scenarios to the time domain[4]. However, achieving reversible tuning, in particular in the visible, remains challenging. The reason is that the metaatoms that constitute the building blocks of phase-gradient metasurfaces are typically made from materials with fixed properties. Indirect tuning was achieved by varying the distance between metaatoms using stretchable substrates[7], by varying the environment that surround metasurfaces[8,9], or modifying metaatoms chemically by exposure to gases or liquids[6,10]. An alternative approach would employ metaatoms that are directly electrically tunable, opening for metaoptics with advanced dynamic control of wavefronts through pixel addressability. In the infrared, electrically tunable plasmonic metaatoms were demonstrated based on the redox-tunable carrier density of conducting polymers[5,11–13]. However, the plasmonic properties of conducting polymers are limited to the infrared and not applicable to visible frequencies due to limited charge carrier density[14,15].

Here, we present a metaatom for visible light that is fully electrically tunable. It also utilizes the redox-tunable properties of a conducting polymer, but with a fundamentally different mechanism which allows it to operate in the visible. Instead of relying on plasmonic properties in the doped high-conducting state, this metaatom is turned on by dedoping the conducting polymer to its neutral state. This introduces an excitonic absorption band in the visible, which appears as a peak in the imaginary permittivity and a dip-peak feature in the real permittivity. Our concept utilizes depolarization of anisotropic nanostructures combined with resonance effects to spectrally split this optical feature into a polarization-dependent response, producing a metaatom capable of geometric phase control in Pancharatnam-Berry (PB) phase metasurfaces[16]. The concept particularly leverages that the dip in the real permittivity can attain negative values if the polymer provides strong resonator strength and low background permittivity. This makes the polymer optically metallic in a selective spectral region such that anisotropic nanostructures can support geometry-dependent exciton-polariton resonances. The effect resembles that of nanostructures made from strongly absorbing dyes[17–20], but with a key difference of being dynamically tunable. The excitonic transition is only active in the neutral state of the polymer and can be turned off by introducing charges by doping. This eliminates the absorption band and turns off the metaatom, which can be repeatedly tuned or gradually varied by low electrical biases (±1 V). This enables dynamically tunable phase-gradient metasurfaces for visible light, which we exemplify for both tunable beam deflection and rewritable holograms.

**Results and discussion**

We validate the concept using poly(3,4-ethylenedioxythiophene) (PEDOT) deposited by vapor phase polymerization (VPP, with OTf as counterion, see Methods). PEDOT provides a neutral absorption band around 630 nm, associated with its distinct blue color. Deposition via VPP further provides uniaxial anisotropy[11] which increases the in-plane oscillator strength through alignment of the dipole transition moments. This amplifies the permittivity dip induced by the absorption band, thereby increasing the likelihood of overcoming the non-resonant permittivity background to induce optically metallic properties. Indeed, *in-situ* spectroscopic ellipsometry confirms that electrochemical dedoping of VPP PEDOT (-1 V, on ITO-coated glass) can induce a spectral region with negative in-plane real permittivity (Fig. 1, a and b; anisotropic results are presented in Extended Data Fig. 1 and Extended Data Fig. 2), down to values of around -2. This means that the dedoped polymer acts optically as a metal in this spectral region, stemming from the excitonic transition instead of mobile charge carriers. Re-doping the material by applying a small positive voltage (+0.5 V) suppresses the absorption band and transitions the material to a transparent dielectric state in the same spectral region (while it instead becomes optically metallic at longer wavelengths as explored for infrared plasmonics[6,13,15,21,22]). Hence, the polymer thin film can be electrically tuned between a state characterized by a pronounced absorption peak that induces optically metallic behavior and a state where it is dielectric and mainly in the visible.

To examine whether the permittivity dispersion of the dedoped VPP PEDOT can facilitate exciton-polariton resonances and depolarization-induced optical anisotropy, we calculate the optical response of ellipsoidal nanostructures (see Methods). This geometry allows us to analytically determine the depolarization factors $L_{x,y,z}$ [23], which determine the polarization-dependence of the quasistatic polarizability through:

$$\alpha_{x,y,z} = V\frac{\varepsilon_{nr}-\varepsilon_s}{\varepsilon_s+L_{x,y,z}(\varepsilon_{nr}-\varepsilon_s)} \quad \text{(Eq. 1)}$$

Here, $V$ is the volume of the ellipsoid, $\varepsilon_{nr}$ is its permittivity (*i.e.* the in-plane permittivity of the polymer) and $\varepsilon_s$ is the surrounding permittivity. Equation 1 gives the quasistatic resonance condition as $L_{x,y,z} = \mathrm{Re}(\varepsilon_s/[\varepsilon_s - \varepsilon_{nr}])$ (Eq. 2). $L_{x,y,z}$ for a nanoellipsoid with semiaxes $a_{x,y,z}$ are then given by[23]:

$$L_{x,y,z} = \frac{a_x a_y a_z}{2}\int_0^\infty \left[\left(s+a_{x,y,z}^2\right)\sqrt{\left(s+a_x^2\right)+\left(s+a_y^2\right)+\left(s+a_z^2\right)}\right]^{-1} \mathrm{ds} \quad \text{(Eq. 3)}$$

A key point here is that $L_{x,y,z}$ is different along different directions for anisotropic ellipsoids (Fig. 1C). This forms the basis for achieving polarization-dependent optical response via exciton-polaritons and depolarization-induced peak shifts. For our polymer with real permittivity values down to around -2, and using $\varepsilon_s = 2$ to represent metaatoms on a substrate in an electrolyte, $L$ needs to be around 0.5 or larger to fulfil Eq. 2. This can be met by the transverse mode of anisotropic ellipsoids excited with light polarized along their width, while the same nanostructure will not meet this criterion for excitation along its length (longitudinal polarization) (Fig. 1c and Extended Data Fig. 3). Figure 1d exemplifies the resulting polarization-dependent optical extinction calculated for a polymer ellipsoid with length $l$ = 200 nm, width $w$ = 50 nm and height $h$ = 75 nm. Transverse excitation leads to an extinction peak located within the optically metallic permittivity region at 590 nm, significantly blueshifted from the material absorption peak (top subpanel). By contrast, excitation along the length of the nanobar (longitudinal polarization) exhibits an extinction peak positioned close to or slightly redshifted from the band absorption. In the doped state of the conducting polymer, the extinction peaks essentially vanish for both polarizations. The calculations support the idea that the dedoped polymeric nanostructure sustains exciton-polariton resonance and depolarization-induced blueshift under transverse polarization, and also that this behavior is not provided under longitudinal polarization. This gives an anisotropic optical response that makes the nanostructure attractive as a metaatom, in particular because its response can be switched off through doping.

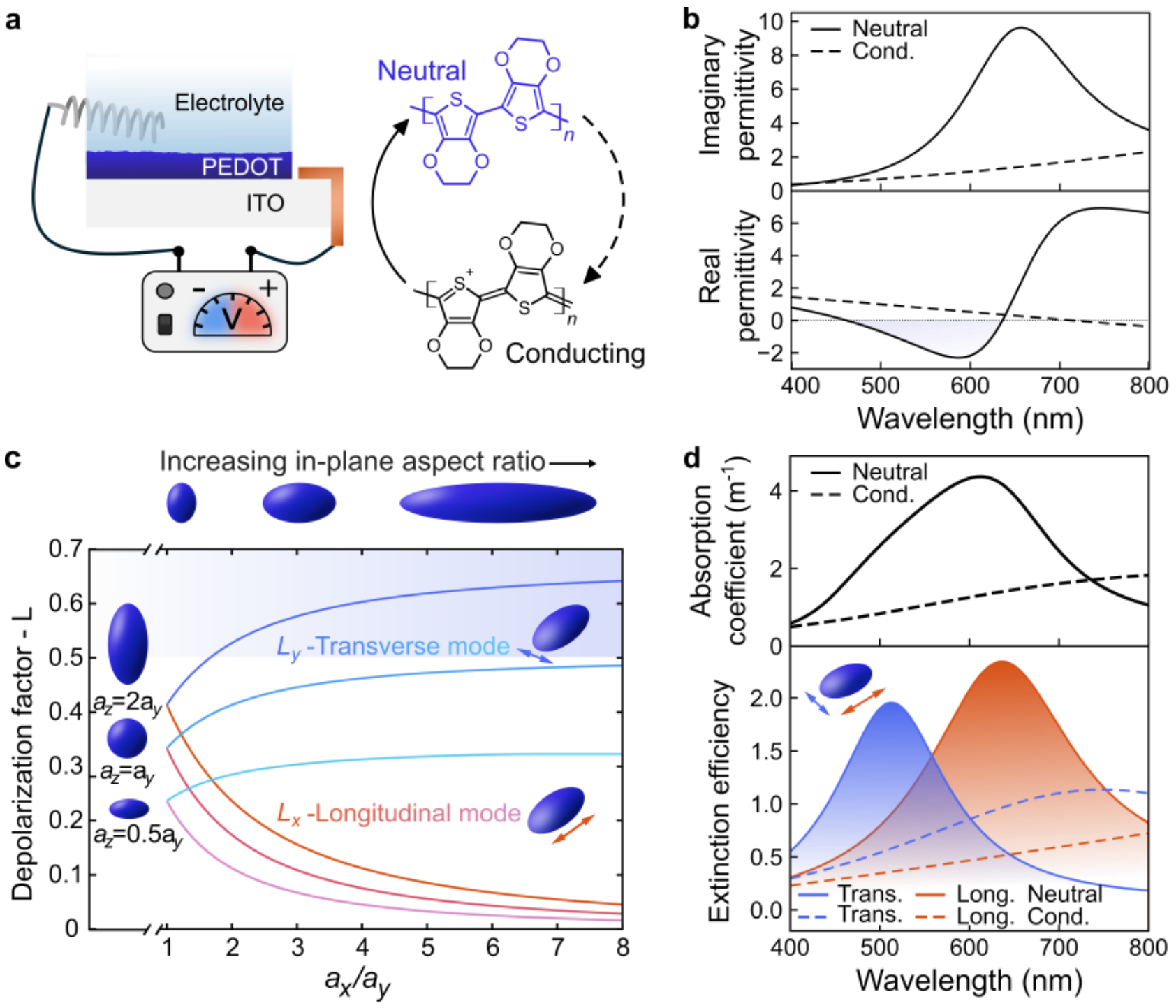

**Fig. 1. Electrically tunable visible metallicity of VPP PEDOT and possibility for exciton-polariton resonances. (a)** Electrochemical tuning process and structure of VPP PEDOT in the conducting and neutral states. **(b)** Measured permittivity of VPP PEDOT in its electrochemically dedoped neutral state (solid lines, -1 V) and doped conducting state (dashed lines, 0.5 V). **(c)** Depolarization factor for ellipsoids of different aspect ratios in their transverse direction ($L_y$, blue shaded lines) and longitudinal ($L_x$, red-shaded lines) directions. **(d)** Calculated absorption coefficient for VPP PEDOT (top panel) and extinction efficiency of a PEDOT nanoellipsoid (bottom panel, $l$ = 200 nm, $w$ = 50 nm and $h$ = 75 nm, $\varepsilon_s = 2$), illuminated by light with transverse polarization (blue) or longitudinal polarization (red). Solid lines represent the neutral dedoped on-state of the polymer and dashed lines show results for the doped conducting off-state.

We experimentally confirm the concept through arrays of anisotropic PEDOT nanobars prepared by electron beam lithography (EBL) on ITO-coated glass (Extended Data Fig. 4). The response for a nonpatterned PEDOT film (70 nm thick, Extended Data Fig. 5), is provided as reference (Fig. 2a). Figure 2c presents the optical response for nanobars of different dimensions, that were electrochemically switched between their dedoped on-state and doped off-state. The insets show scanning electron microscopy (SEM) images (complemented by additional SEM images in Extended Data Fig. 6, and AFM images in Extended Data Fig. 5). In the doped conducting off-state (+0.5 V), the nanobars show effectively no optical response in the visible regardless of size and polarization, which agrees with the suppressed absorption band and non-metallic visible permittivity of the polymer in this state. Changing the electrical bias to -1 V dedopes the polymer to its neutral state, which turns on the excitonic transition and the depolarization-controlled anisotropic response. In accordance with the calculated nanoellipsoid example above, transverse excitation shows extinction peaks that are blueshifted from the thin film and material absorption. This finding verifies that the optical response is governed by the nanostructuring, in line with exciton-polariton resonance and depolarization effects. Further support is obtained through a systematic variation in peak position with nanobar dimension, with increasing blueshift for nanobars with decreasing width and increasing in-plane aspect ratio. This leads to larger $L$ for the transverse mode, such that the resonance condition (Eq. 2) can be fulfilled at a lower magnitude of $\varepsilon_{nr}$ which occurs at shorter wavelengths (Fig. 1c). Optical calculations further show that the blueshift does not suddenly disappear for systems not strictly meeting the resonance condition. Instead, blueshifted peaks can still be provided by the large depolarization factor combined with the material absorption peak, further expanding the application scenario of the concept (Extended Data Fig. 7 and Extended Data Fig. 8). Optical simulations (Fig. 2, b and d) support the experimental findings, with minor differences attributed to detailed geometries, uniformity and permittivity. The optical response at longitudinal polarization is clearly different from transverse polarization for all nanobar dimensions, with the dedoped extinction peaks not being blueshifted but instead close to the thin film response (Fig. 2, c and d).

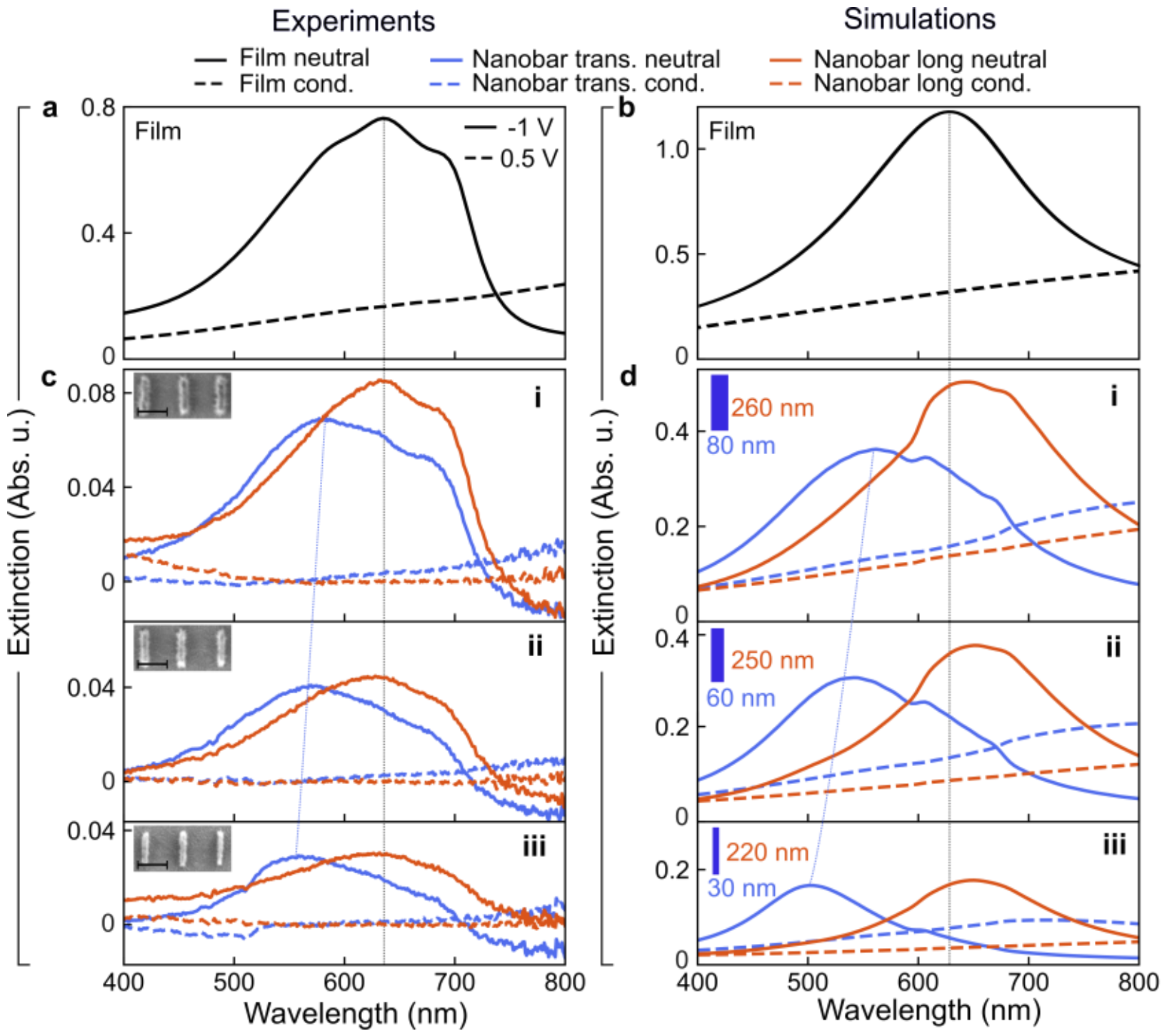

**Fig. 2. Size-dependent exciton-polariton resonances in neutral PEDOT nanobar antenna arrays.** **(a)** Experimental extinction of a PEDOT film (≈70 nm thick, on ITO-coated glass) electrochemically switched to its dedoped neutral state (solid line, -1 V) or doped conducting state (dashed line, +0.5 V). **(b)** Simulated extinction spectra for the same system as in (A). **(c)** Experimental extinction of PEDOT nanobar arrays of different dimensions (≈70 nm thick on ITO-coated glass) in their electrochemically dedoped (solid lines, -1 V) and doped (dashed lines, +0.5 V) states, with decreasing nanobar width from top (i) to bottom (iii). The insets are SEM images of respective nanobar arrays, with 200 nm scale bars. **(d)** Simulated extinction results for PEDOT nanobar arrays on ITO-coated glass with dimensions corresponding to the experimental arrays of panel (70 nm thick) (C). The periodicity was 250 nm along *x* (transverse) and 500 nm along *y* (longitudinal). In panels (C) and (D), the vertical black dotted lines mark the film extinction peak position and the blue (non-vertical) dotted lines serve as visual guides of the blueshift of the transverse exciton-polariton peak for different antenna dimensions.

We demonstrate that the polarization-dependent nanobars can be used as tunable visible metaatoms in PB phase-gradient metasurfaces[16]. PB metasurfaces provide control of circularly polarized light (CPL) through spin-reversed scattering of anisotropic metaatoms, where each metaatom contributes with a local phase shift $\varphi$ that can be controlled from 0 to $2\pi$ through their rotation angle $\theta$ as $\varphi = 2\theta$ [24]. This enables the design of arbitrary phase maps and metasurfaces with functions ranging from beam deflection and lensing to holograms and advanced polarization control[3]. Optical simulations confirm that the excitonic PEDOT nanobars can provide such continuous and predictable 0-$2\pi$ phase coverage (Extended Data Fig. 9), supporting their applicability for active PB metasurfaces. We test this by constructing a beam-deflecting metasurface with a linearly varying phase along $x$ and fixed phase along $y$ (schematically depicted in Fig. 3a). This produces a spin-reversed beam that is deflected along the $x$ axis, as illustrated in Figure 3b which presents the simulated phase propagation for left-handed circular polarized (LCP) light upon excitation at normal incidence with right-handed circular polarized (RCP) light (at 632 nm). We focus here on operation at 632 nm at which the optical anisotropy is large, while noting that the principle works also at other wavelengths (Extended Data Fig. 10). Intensity simulations in Fig. 3c confirm that the dedoped on-state of the metaatoms produces a spin-reversed beam deflected at an angle of 8°, in accordance with the design based on 4.5 μm wide supercells (see Methods for details). Importantly, this deflected beam essentially disappears for the conducting off-state of the nanoantennas (Fig. 3d). Figure 3e and 3f present experimental results for the equivalent metasurface, illuminated by RCP light at 632 nm at normal incidence and with the doping level of the metasurface varied electrochemically (indicated by the downward time arrow) in a fluidic cell. The results confirm the simulated predictions and show a clear deflected beam of opposite handedness already when the nanoantennas are partially dedoped (-0.5 V bias, i). Further dedoping of the nanoantennas (-1 V, ii) increases the deflected beam intensity. By contrast, doping the metasurface by stepwise increasing the applied bias potential decreases the deflected beam intensity (0 V, iii) and turns it off (0.5 V, iv). In this off-state, primarily non-deflected light that was not sufficiently suppressed by the polarization optics could be detected (see experimental setup in Extended Data Fig. 11). Importantly, the tuning process is reversible, providing full dynamic modulation at low bias potentials. This is shown through Figure 3g, presenting the variation in deflected beam power upon repeatedly switching a metasurface between its on-state (-1 V) and off-state (+0.5 V). The enlarged part presents data for one cycle together with extracted switching times for 10%-90% (black) and 20%-80% (grey) of the total switching, showing on-switching (dedoping) on the order of a second (0.7 s-1.5 s) and off-switching (doping) at around 0.13 s-0.25 s. Extended Data Figure 12 provides additional results upon stepwise tuning, and Extended Data Figure 13 experimentally confirms that the mechanism works also at a different wavelength (589 nm).

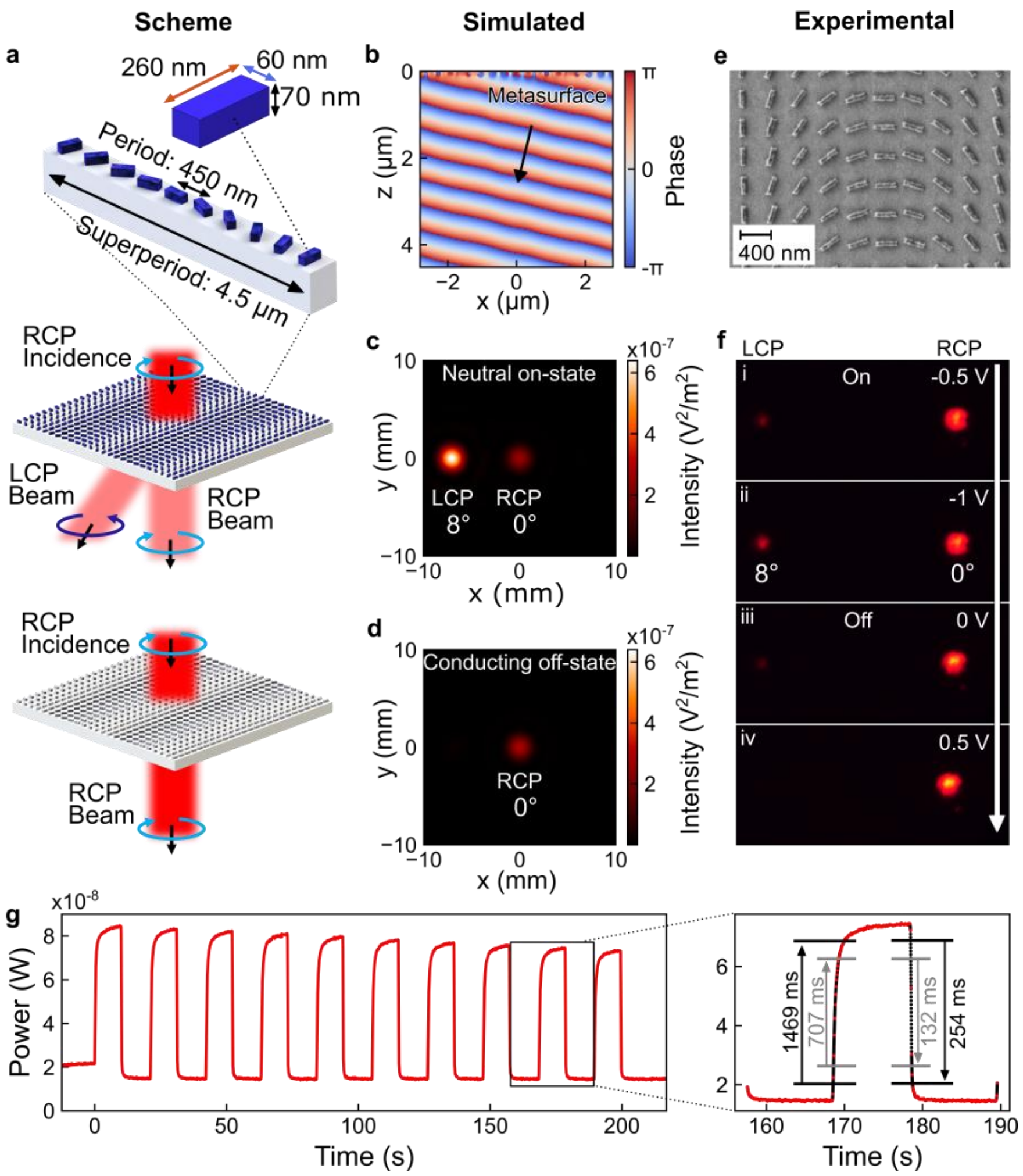

**Fig. 3. Electrically tunable visible phase-gradient metasurface. (a)** Schematic illustrations of metasurface designed for dynamically tunable beam deflection using repeating supercells of 10 excitonic metaatoms with gradually varying orientation. **(b)** Simulated LCP phase propagation for the metasurface in the neutral on-state of the nanoantennas, excited by normal incidence RCP light from below. The outgoing wavefront is deflected by 8°. **(c)** Simulated intensity distribution 50 mm from the metasurface in the neutral on-state after illumination by a RCP beam of finite size, showing both the spin-reversed deflected LCP beam and also the RCP light (with the latter reduced by a factor 350 to account for attenuation by the polarization optics in the experimental setup). **(d)** Same as in panel (c) but for the conducting off-state of the metasurface. **(e)** SEM image of a produced beam deflecting metasurface with same design as in the scheme and simulations. **(f)** Experimental beam deflection results for such metasurface, with doping level controlled electrochemically in steps (i) -0.5 V, (ii) -1 V, (iii) 0 V, and (iv) 0.5 V. **(g)** Measured power of deflected beam upon repeated switching between -1 V (on-state) and 0.5 V (off-state), with rightmost part providing switching time data for one cycle. Both simulations and experiments operated at 632 nm wavelength.

Finally, we employ the dynamic metaatom for more complex wavefront control in the form of a dynamic metahologram. We designed a star-shaped hologram using the Gerchberg–Saxton algorithm (Fig. 4a). This approach allows the calculation of 2D phase pattern (Fig. 4b) that can be repeated in *x* and *y* as supercells to a larger metasurface, suppressing laser speckle and improving the fidelity of the final holographic image (Extended Data Fig. 14). The simulation in Figure 4d shows how two-by-two such supercells, constructed by dedoped PEDOT nanobars, create a spin-reversed dispersive holographic image which is not visible in the doped off-state (Fig. 4e).

Experimentally, the holographic phase map containing multiple supercells (10 by 10) was imposed by an additional parabolic phase modulation to focus the hologram at a distance 5 mm from the metasurface (Fig. 4, c and f).Figure 4g present experimental result for the metahologram in the excitonic on-state, reproducing well the desired holographic image (see Extended Data Fig. 15 for experimental details). Turning the metaatoms to their off-state by doping erases the hologram image (Fig. 4h). The metahologram can then be electrochemically turned on again, gradually controlled to any desired intermediate state.

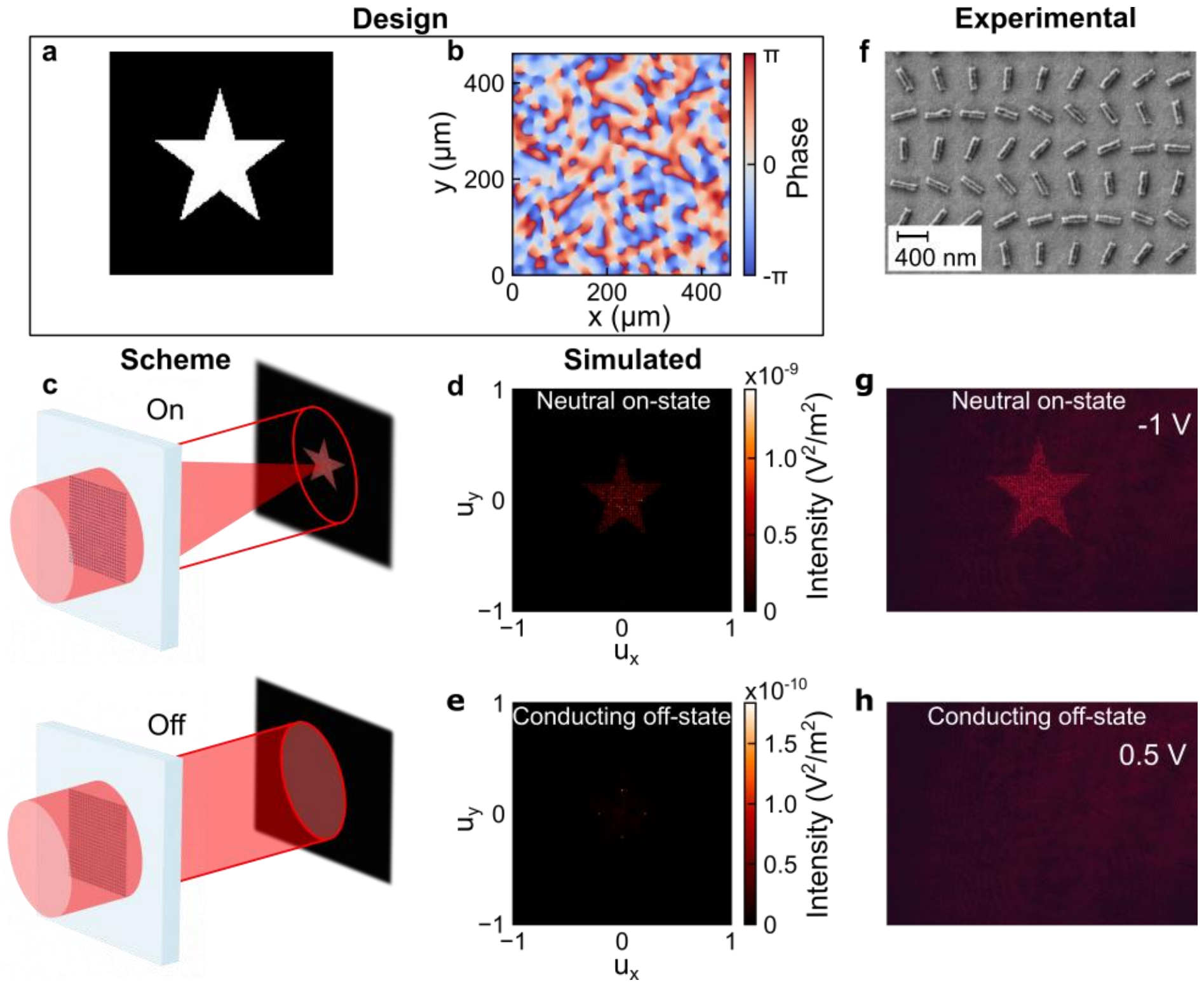

**Fig. 4. Dynamically tunable visible metahologram. (a)** Desired hologram image. **(b)** Calculated phase map to reproduce the star-shaped image. **(c)** Illustration of the metasurface producing a focused hologram in the neutral on-state (top) which is turned off in the conducting doped state of the metasurface (bottom). **(d)** Simulated spin-reversed response of a two-by-two supercell hologram in the neutral on-state of the PEDOT metaatoms. Here, $u_x = \sin(\theta)\cos(\varphi)$ and $u_y = \sin(\theta)\sin(\varphi)$, where $\theta$ is the polar incidence angle relative to the surface normal, $\varphi$ is the in-plane azimuthal angle, and $u$ is the unit wave-propagation direction. Intensity values correspond to 1 m distance from the metasurface. **(e)** Same as in panel (d) but for the conducting off-state of the metasurface. **(f)** SEM image of part of a metahologram device. **(g)** Photograph of hologram image produced by the final metasurface (10 by 10 supercells and lens function) in its dedoped on-state (-1 V). **(h)** Equivalent results to that in panel (g) but after applying a 0.5 V bias to turn the metasurface off by doping.

## Conclusions

Our study demonstrates an electrically tunable metaatom for visible light and its use for active phase-gradient metasurfaces. We used PEDOT as model material, but the concept is generic and applicable also for other materials with dynamically tunable absorption band and low background permittivity[25]. Likewise, the reported principle is not limited to electrical control but can be extended to other tuning modalities or adaptive functionality using materials responsive to light, heat, or even biological or chemical triggers. For example, the PEDOT material studied here can be tuned also *via* chemical stimuli[6]. New materials may further be specifically designed to provide desired modality and properties suitable for targeting specific spectral ranges of operation. Combined with pixel addressability, our study opens for digitated visible metaoptics with complete dynamic control of optical wavefronts in the visible.

## Methods

### Fabrication of PEDOT films

PEDOT films were made by vapor phase polymerization of 3,4-ethylenedioxythiophene (EDOT), using trifluoromethanesulfonate (OTf), as a counterion. Glass substrates with a 50 nm layer of indium tin oxide (ITO, Biotain) were washed by sonicating subsequently in an aqueous solution with 2% Hellmanex, DI water, acetone, and isopropanol for 10 minutes each. The substrates were oxygen plasma cleaned for 5 minutes at 100 W. An oxidant solution containing 3 wt.% iron(III) OTf (Fisher Scientific), 20 wt.% Poly(ethylene glycol)-*block*-poly(propylene glycol)-*block*-poly(ethylene glycol) (PEG-PPG-PEG, Sigma Aldrich), and 77 wt.% ethanol (Solveco) was spin coated on the substrates (1500 r.p.m., 60 s). The substrates were baked for 30 s at 70 °C. EDOT (Sigma-Aldrich) was evaporated onto the oxidant-coated substrates in a vacuum chamber (-1 bar) at 60 °C for 30 minutes. The VPP reaction was continued for 1 minute in ambient conditions while heating the substrates at 60 °C. The samples were then washed in ethanol for four hours, followed by drying with nitrogen.

### Fabrication of nanobars

Electron beam lithography (EBL, Voyager (50kV), Raith) was used to pattern the PEDOT films. For the nanoantenna arrays, a 500 nm layer of the positive EBL resist ZEP520A (Zeon), was spin coated onto the PEDOT film (2000 r.p.m., 60 s). For the phase-gradient metasurfaces, a 200 nm layer of ZEP520A (diluted 1:1 in anisole), was spin coated onto the PEDOT film (1500 r.p.m., 60 s). The resist was then patterned according to the intended nanopattern design and developed in n-amyl acetate (ZED-N50, Zeon) for 80s. Reactive ion etching (RIE) was used to transfer to pattern from the resist to the PEDOT film (50 W). For the nanoantenna arrays, the RIE time was varied from 100 s to 180 s to create different sizes of nanobars. For the phase-gradient metasurfaces, the RIE time was 80s. The remaining resist was removed by soaking the samples in AR 300-72 (Allresist) for 1 hour.

### Structural characterization

Thicknesses of polymer films were determined by profilometry (Dektak XT, Bruker). Thicknesses of nanobars were determined by atomic force microscopy (AFM, Dimension 3100, Veeco), while lateral dimensions were determined by scanning electron microscopy (SEM, Sigma 500, Zeiss).

**In-situ spectroscopy**

Constant potentials of -1 V and +0.5 V were applied to the films and nanobars to switch between their neutral and conducting state, respectively. To ensure the stabilization of the current, potentials were applied for at least 100 s before a measurement. Additionally, five cycles of cyclic voltammetry (at 0.5 V/s) were done before performing a measurement on a pristine sample to exclude the initial response. Acetonitrile (Sigma Aldrich) containing 0.05 M tetrabutylammonium tetrafluoroborate (TBA $BF_4$, Sigma Aldrich) was used as an electrolyte. Extinction spectra were measured using a UV-vis spectrometer (Lambda a1050+, Perkin Elmer Instruments) and a spectroelectrochemical cell (redox.me). A two-electrode setup was used, with the film or nanobar sample as working electrode and a platinum wire counter electrode. A measurement of a blank ITO substrate in the spectroelectrochemical cell was used as reference. Reflective ellipsometry was performed with an RC2 spectroscopic ellipsometer (J. A. Woollam Co.) and an ellipsometry electrochemical cell with an angle of incidence of 70 degree (redox.me). For ellipsometry, a three-electrode setup was used, with the film or nanobar sample as working electrode, a platinum wire counter electrode and a Ag/Ag+ reference electrode. The CompleteEASE software was used to fit the data with a Drude-Lorentz model.

**Optical calculations**

The optical extinction $E$ of ellipsoidal nanostructures was calculated as $E = k\mathrm{Im}(\alpha')$, where $\alpha'$ is the radiation-corrected dipolar polarizability described by:

$$\alpha' = \alpha \left[1 - i\frac{k^3}{6\pi}\alpha\right]^{-1}$$

, with $\alpha$ being the quasistatic polarizability determined by Eq. 1 using the in-plane permittivity. The polymer absorption coefficients $\alpha_{\mathrm{material}}$ were determined by:

$$\alpha_{\mathrm{material}} = \frac{4\pi k}{\lambda}\mathrm{Im}(\tilde{n})$$

where $\tilde{n}$ is the complex refractive index of the polymer derived from the in-plane complex permittivity. Extinction efficiency is defined as extinction cross section divided by the cross-sectional geometrical area of the same nanostructure.

**Finite-Difference Time-Domain (FDTD) Simulations**

FDTD simulations were performed to quantitatively analyze the response of single nanostructures, periodic arrays and phase-gradient metasurfaces. The nanobars were defined according to the geometrical dimensions extracted from SEM characterization and placed on a substrate. The anisotropic permittivity dispersions of PEDOT in both the doped and dedoped states were implemented from spectroscopic ellipsometry.

For simulations of individual nanostructures, a single nanobar was positioned in a three-dimensional rectangular simulation domain terminated by perfectly matched layers (PML) in all directions. A broadband total-field scattered-field (TFSF) plane-wave source illuminated the structure, with linear polarization either parallel or perpendicular to the long axis of the bar to selectively excite longitudinal and transverse modes. A nonuniform conformal mesh was employed, with a local mesh override of 2–3 nm surrounding the nanoantenna to accurately resolve near-field gradients and geometric edges, while coarser meshing was applied in regions far from the antenna. Frequency-domain field monitors were implemented as closed two-dimensional box surfaces enclosing the antenna. A box monitor placed outside the TFSF source was used to collect the scattered power, while a second box monitor tightly enclosing the nanoantenna inside the TFSF source was used to determine absorption. The extinction spectra were obtained by summing the scattering and absorption components.

A metasurface consisting of nanobar arrays was simulated in a similar manner to single nanostructures, with periodic boundary conditions applied in the $x$ and $y$ directions to represent an infinite array. The structure was illuminated by a normally incident plane-wave source, and two-dimensional frequency-domain field monitors were used to record the transmitted fields. The plane-wave polarization was rotated to obtain the transverse and longitudinal mode spectra. The extinction spectrum of the metasurface was determined from the reduction of transmitted intensity relative to the incident power, accounting for the combined effects of scattering and absorption.

To evaluate the dependence of phase shift on the nanobar rotation angle, periodic arrays of nanobars were simulated with periodic boundary conditions applied in the $x$ and $y$ directions to represent an infinite metasurface. The nanobar within the unit cell was systematically rotated while keeping all other geometrical parameters fixed. The structure was illuminated at normal incidence with circularly polarized light, and the transmitted fields were recorded using a frequency-domain monitor placed above the metasurface. The transmitted field was decomposed into circular polarization components to isolate the spin-reversed (cross-polarized) contribution. The phase was extracted from the complex cross-polarized transmission coefficient, and its variation was analyzed as a function of the nanobar rotation angle, revealing the rotation-dependent phase shift.

To analyze the phase propagation of phase-gradient metasurfaces, simulations were performed using supercells containing several nanobars arranged according to the designed rotation angles of the phase-gradient profile. Periodic boundary conditions were applied in the x and y directions to represent an infinitely repeated structure. The metasurfaces were illuminated at normal incidence using a circularly polarized plane-wave source. Two-dimensional frequency-domain field monitor placed across a supercell was used to simulate and extract the phase propagation through the metasurface from the complex field distribution, enabling analysis of the spatial phase variation introduced by the phase gradient.

To simulate spin-reversed deflected beams, far-field projection was performed using a $5 \times 10$ supercell to represent a finite portion of the phase-gradient metasurface. Perfectly matched layers (PML) were applied in all directions to properly model the finite aperture and suppress artificial boundary reflections. The metasurface was illuminated with circularly polarized light, and the transmitted fields were decomposed to isolate the spin-reversed (cross-polarized) component. A two-dimensional field monitor was placed on a closed surface above the phase-gradient metasurface. The recorded field data were projected onto a planar surface with a distance 50 mm from the metasurface using a near-to-far-field transformation to calculate the far-field intensity distribution of the spin-reversed deflected LCP beam and also the RCP light. The deflection angle of the spin-reversed beam was calculated using the tangent relation, defined as the distance betweenthe deflected LCP beam and the RCP beam devided by the projection distance.

**Hologram design and simulations**

The hologram phase profile was designed using the Gerchberg–Saxton algorithm to generate a target star-shaped intensity pattern in the far field. Starting from the desired image, an iterative Fourier transform procedure was employed to retrieve the corresponding phase-only distribution while constraining the amplitude in the metasurface plane. After convergence, a two-dimensional phase map was obtained, representing the required spatial phase modulation for holographic reconstruction. The calculated phase distribution was then discretized and mapped onto the nanobars by assigning their in-plane rotation angles according to the Pancharatnam–Berry (geometric) phase principle. The 128 × 128 phase map was implemented within a supercell configuration, where each nanobar orientation corresponded to the required local phase value.

To evaluate the optical performance of the designed hologram, FDTD simulations were performed on a metasurface composed of 2 × 2 supercells, each containing 128 nanobars. The geometrical parameters of nanobars were extracted from SEM characterization and subsequently implemented in simulation, where the nanobars were placed on a substrate. The anisotropic permittivity of the nanobar material (PEDOT), in both doped and dedoped states, was implemented using spectroscopic ellipsometry data.

A three-dimensional simulation domain was employed with perfectly matched layers (PML) applied in all directions. The metasurface was illuminated at normal incidence using a circularly polarized plane-wave source covering the entire structure, where circularly polarization was implemented by two perpendicular linearly polarized components with a 90° phase offset. A nonuniform mesh was implemented, with a local mesh refinement of 2 – 3 nm around the nanobars to accurately resolve subwavelength features and near-field variations, while a coarser mesh was used away from the structure to reduce computational cost.

To obtain the spin-reversed response, a frequency-domain field monitor was placed on a closed surface above the metasurface. The recorded field data were used to calculate the far-field radiation pattern using hemispherical far-field projection. The far-field projection was plotted in terms of the direction cosines $u_x = \sin(\theta)\cos(\varphi)$ and $u_y = \sin(\theta)\sin(\varphi)$, where θ is the polar angle measured from the surface normal, φ is the in-plane azimuthal angle, and u represents the unit wave-propagation direction.

**Metasurface characterization**

Beam deflection and hologram metasurfaces in a spectroelectrochemical cell were illuminated by circularly polarized laser light. The electrochemical set up was similar as described under in-situ spectroscopy. Laser light sources were a red laser (632 nm, 10 mW, Coherent) and a yellow diode pumped solid state (DPSS) laser (589 nm, 100 mW, Xinland). RCP light was made by passing the laser light beam through a polarizer and a quarter wave plate. After the light has passed through the sample the RCP light is partially filtered out by passing the light through another quarter wave plate and polarizer. For the beam deflection metasurfaces the deflected LCP light and the RCP light that could not be filtered out is projected onto a screen and then captured by a CMOS camera. For the hologram metasurfaces the LCP light that forms the hologram is collected by a 10x objective lens and then passed through a tube lens to a CMOS camera.

**Acknowledgments:**

The authors acknowledge funding from the European Research Council (Consolidator Grant, 101086683),the Knut and Alice Wallenberg Foundation (Wallenberg Academy Fellow, KAW 2019.0163, 2020.0301),the Swedish Research Council (VR, Consolidator Grant, 2020-00287),and the Swedish Government Strategic Research Area in Materials Science on Functional Materials at Linköping University (Faculty Grant SFO-Mat-LiU No. 2009 00971).

**Author contributions:**

M.P.J. conceived the study. J.W., L.L., H.M.S. and S.K. fabricated PEDOT films and nanobars and did structural characterization. L.L. performed in-situ ellipsometry on PEDOT films and fitted the data. J.W. acquired extinction spectra of films and nanobars, with the help of L.L. and S.K.

L.L. and H.M.S. performed FDTD simulations, with the help of D.L. and G.C. M.P.J did optical calculations. H.M.S. designed the metahologram and performed the corresponding simulations. H.M.S and J.W. characterized the beam deflection and hologram metasurfaces. J.W., M.P.J. and H.M.S. wrote the manuscript and all authors reviewed and commented on the manuscript. M.P.J., Y.D. and S.C. supervised the project.

**Competing interests:** The authors declare no competing interest.

**Corresponding author:** Correspondence to Magnus P. Jonsson

## Extended Data

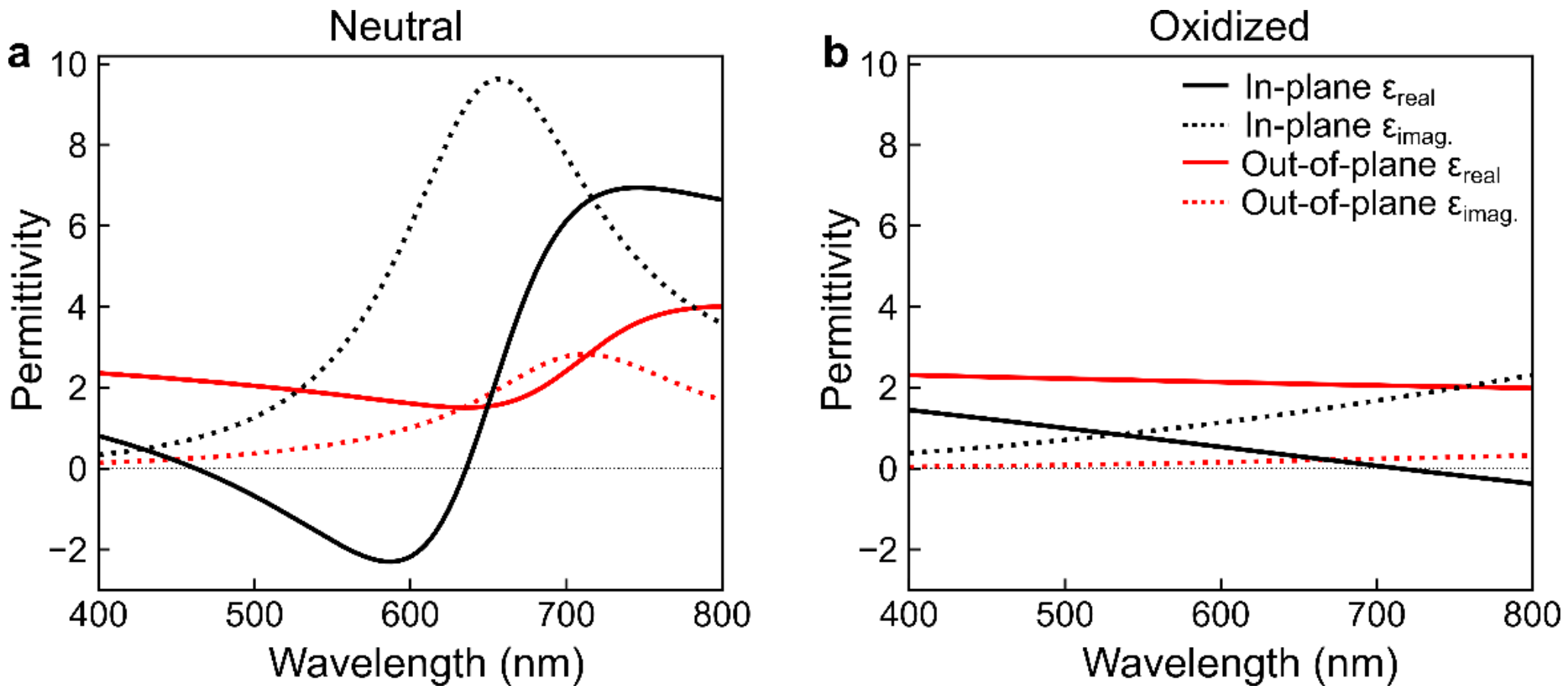

**Extended Data Fig. 1. Anisotropic permittivity of PEDOT in the visible. (a)** Measured permittivity of electrochemically dedoped PEDOT (-1 V). The peak in the in-plane imaginary permittivity signals the excitonic transition. The out-of-plane imaginary permittivity peak is less intense, showing that the PEDOT chains are aligned such that excitonic transition contributes to a larger extent to the in-plane response than to the out-of-plane response. The strength of the in-plane excitonic transition causes a dip in the real permittivity that reaches negative values while the out-of-plane real permittivity does not reach negative values. **(b)** Permittivity of electrochemically doped PEDOT (0.5 V). The excitonic transition is suppressed. Instead, the in-plane real permittivity indicates that material becomes metallic at longer wavelengths.

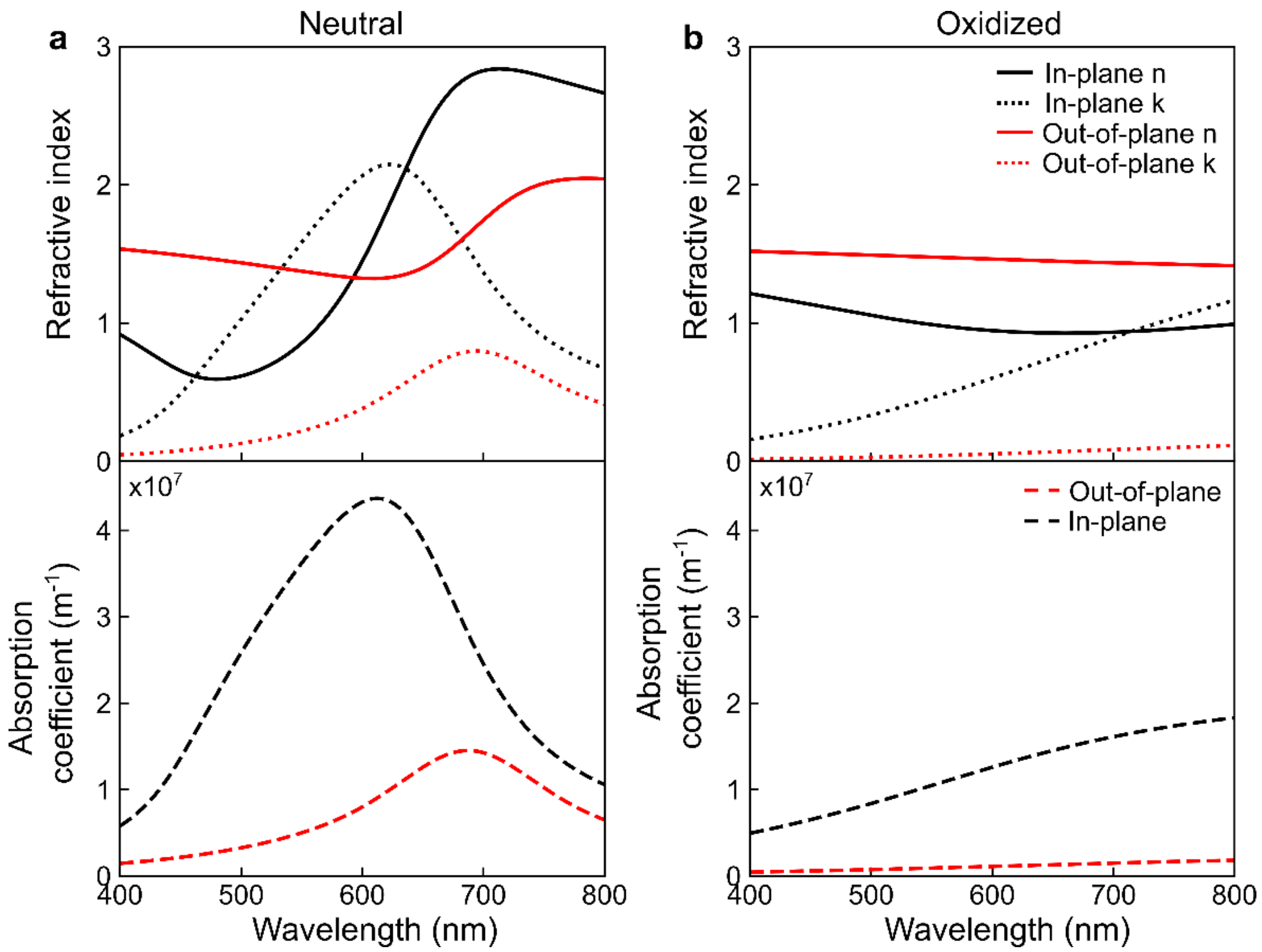

**Extended Data Fig. 2. Anisotropic refractive index and absorption coefficients of PEDOT in the visible** (obtained from the data in Fig. S1)**. (a)** Refractive index (top panel) and absorption coefficient (bottom panel) of dedoped PEDOT. **(b)** Refractive index (top panel) and absorption coefficient (bottom panel) of conducting doped PEDOT.

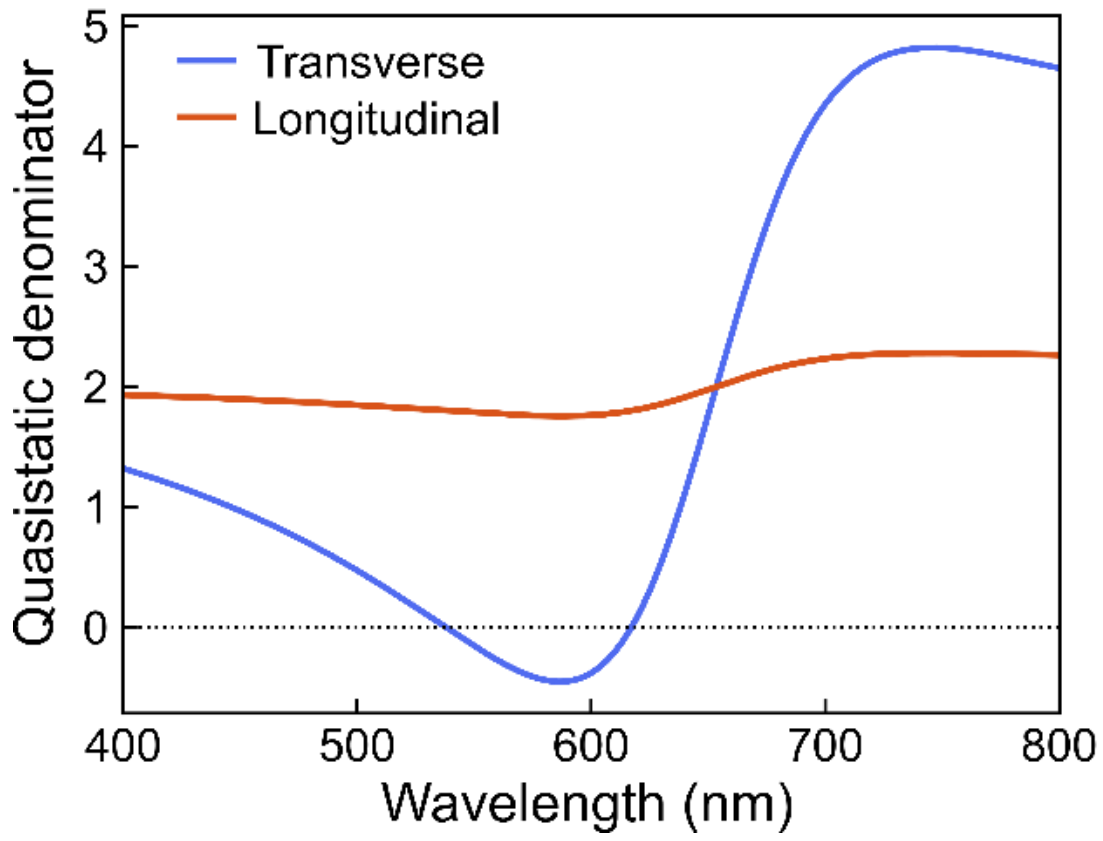

**Extended Data Fig. 3. Quasistatic resonance condition.** Real part of the denominator in Eq. 1 for transverse (blue) and longitudinal (red) polarization for an ellipsoid with dimensions $l$ = 200 nm, $w$ = 50 nm and $h$ = 75 nm. Transverse excitation crosses zero (fulfilling Eq. 2) while longitudinal excitation does not.

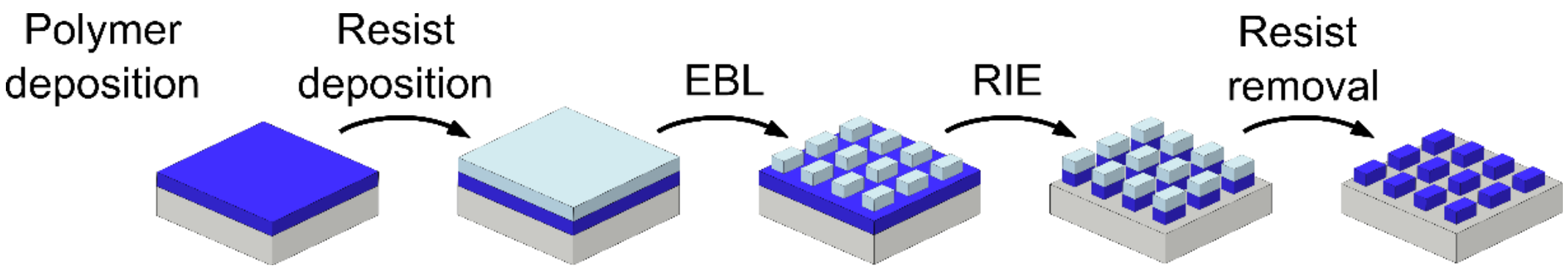

**Extended Data Fig. 4. Fabrication of PEDOT:OTf nanobars.** Schematic illustration showing the main process flow of the nanofabrication of nanobar arrays and metasurfaces.

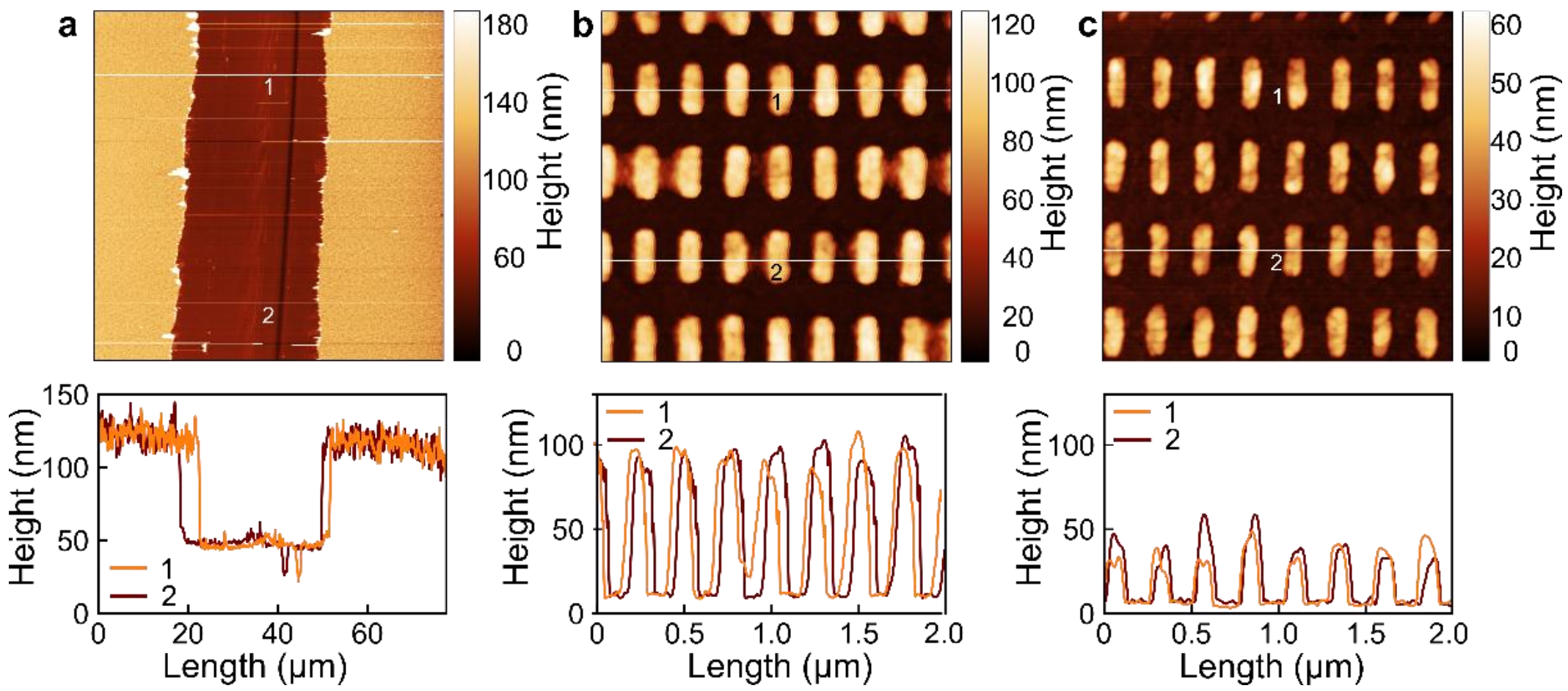

**Extended Data Fig. 5. AFM characterization of PEDOT film and nanobar array. (a)** AFM image of PEDOT thin film, showing film thickness of approximately 70 nm. **(b)** AFM image of a region of a PEDOT nanobar array, showing thickness of approximately 90 nm. **(c)** AFM image acquired from a different region than in (b), showing a lower thickness of approximately 40-60 nm. Spatial non-uniformity may originate from variation in removal of EBL resist after etching or from non-uniform film thickness.

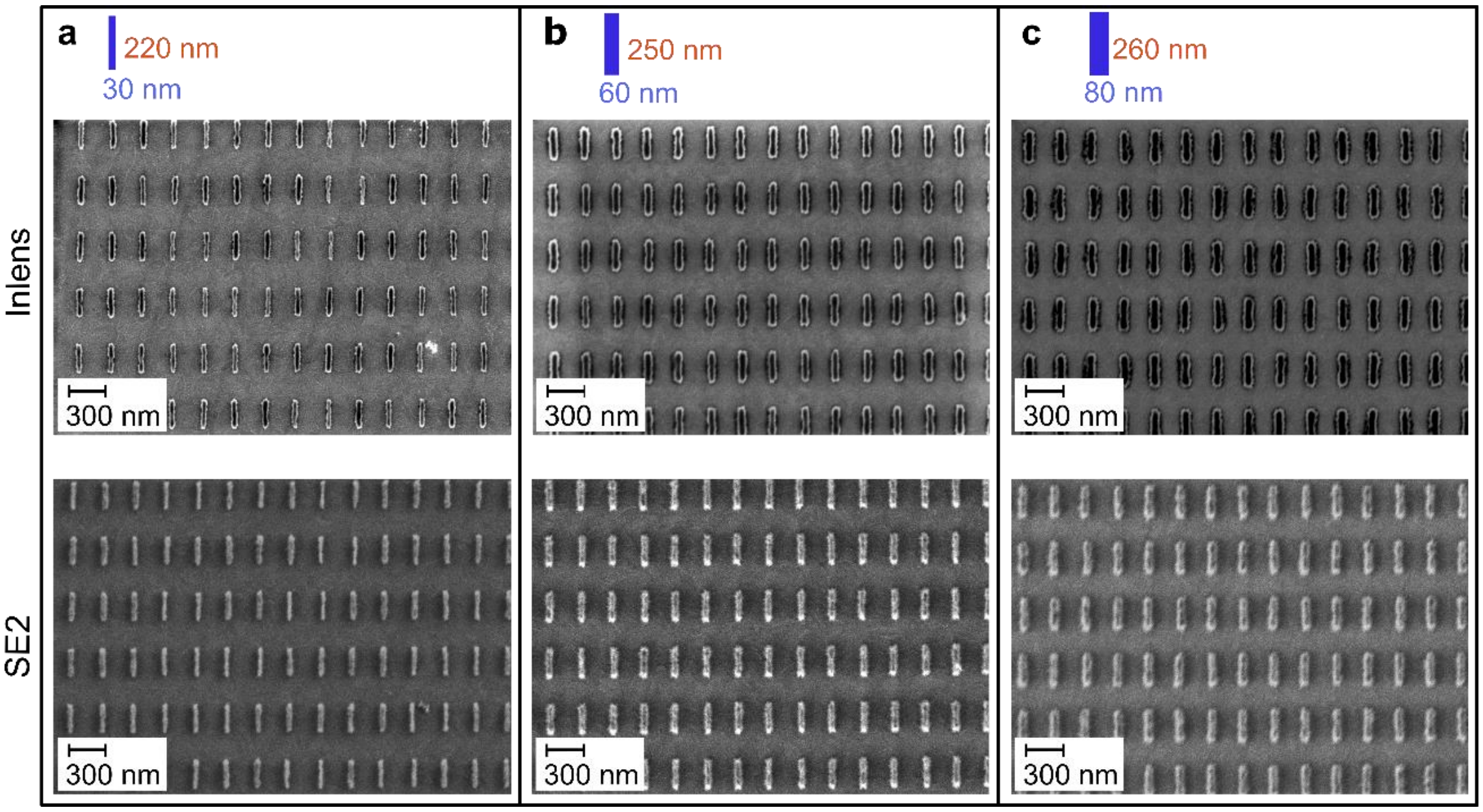

**Extended Data Fig. 6. SEM images of PEDOT nanobar antenna arrays of different sizes. (a)** The smallest nanobars in the series, with dimensions of 30 nm by 220 nm. Some of the nanobars are slightly wider (50 nm by 220 nm). **(b)** Nanobars with dimensions of 60 nm by 250 nm. **(c)** The largest nanobars in the series, with dimensions of 80 nm by 260 nm. Shown are images made with an Inlens detector and with an SE2 detector. For the size determination the SE2 images were used.

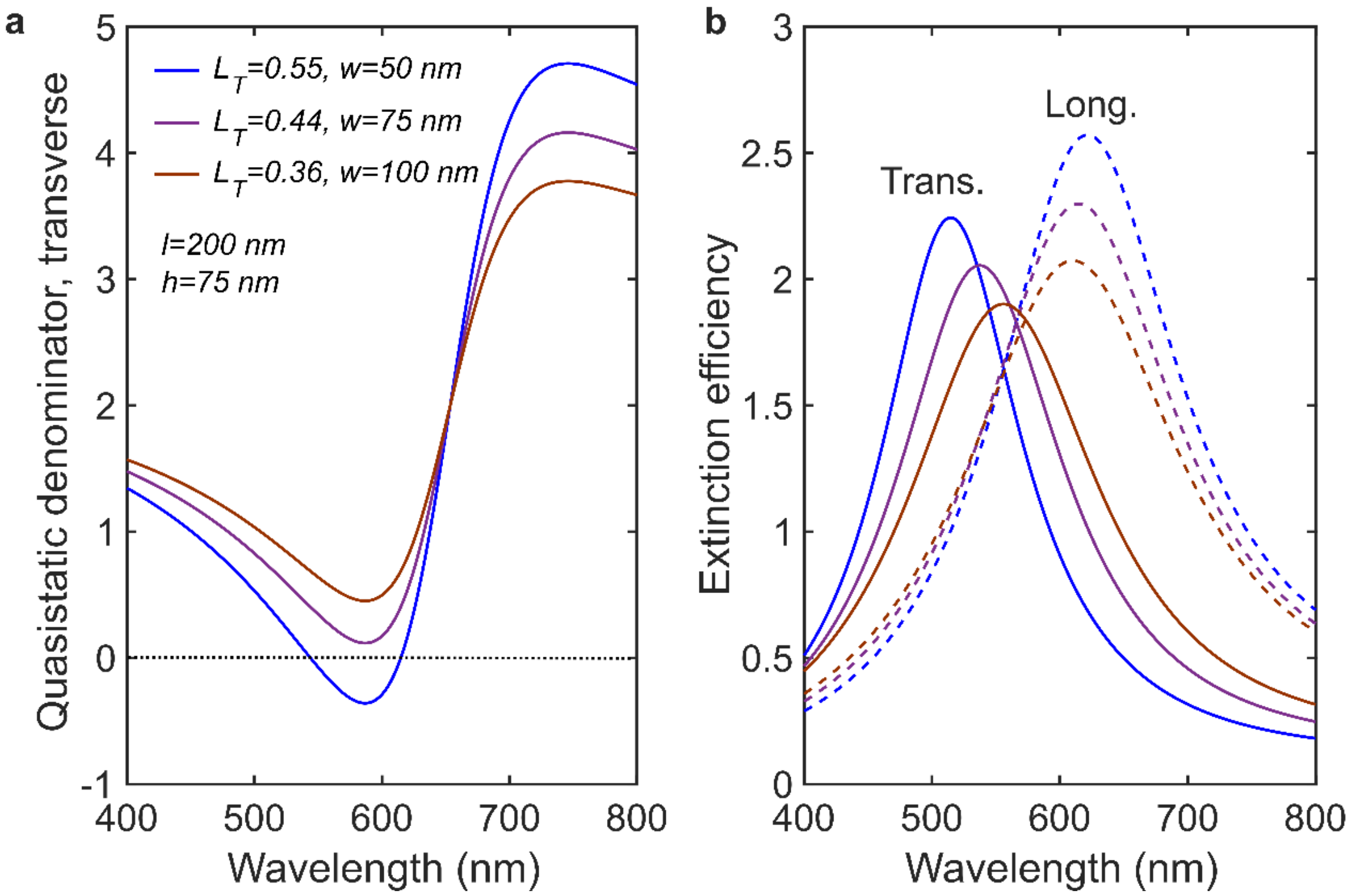

**Extended Data Fig. 7. Moving out of resonance by varying dimension of dedoped PEDOT nanoellipsoids. (a)** Real part of the denominator in Eq. 1 for transverse polarization for ellipsoids with $l$ = 200 nm, $h$ = 75 nm, and different widths and thereby depolarization factors as indicated in the legend. **(b)** Extinction efficiency for the same ellipsoids for transverse (solid lines) and longitudinal (dashed lines) polarization.

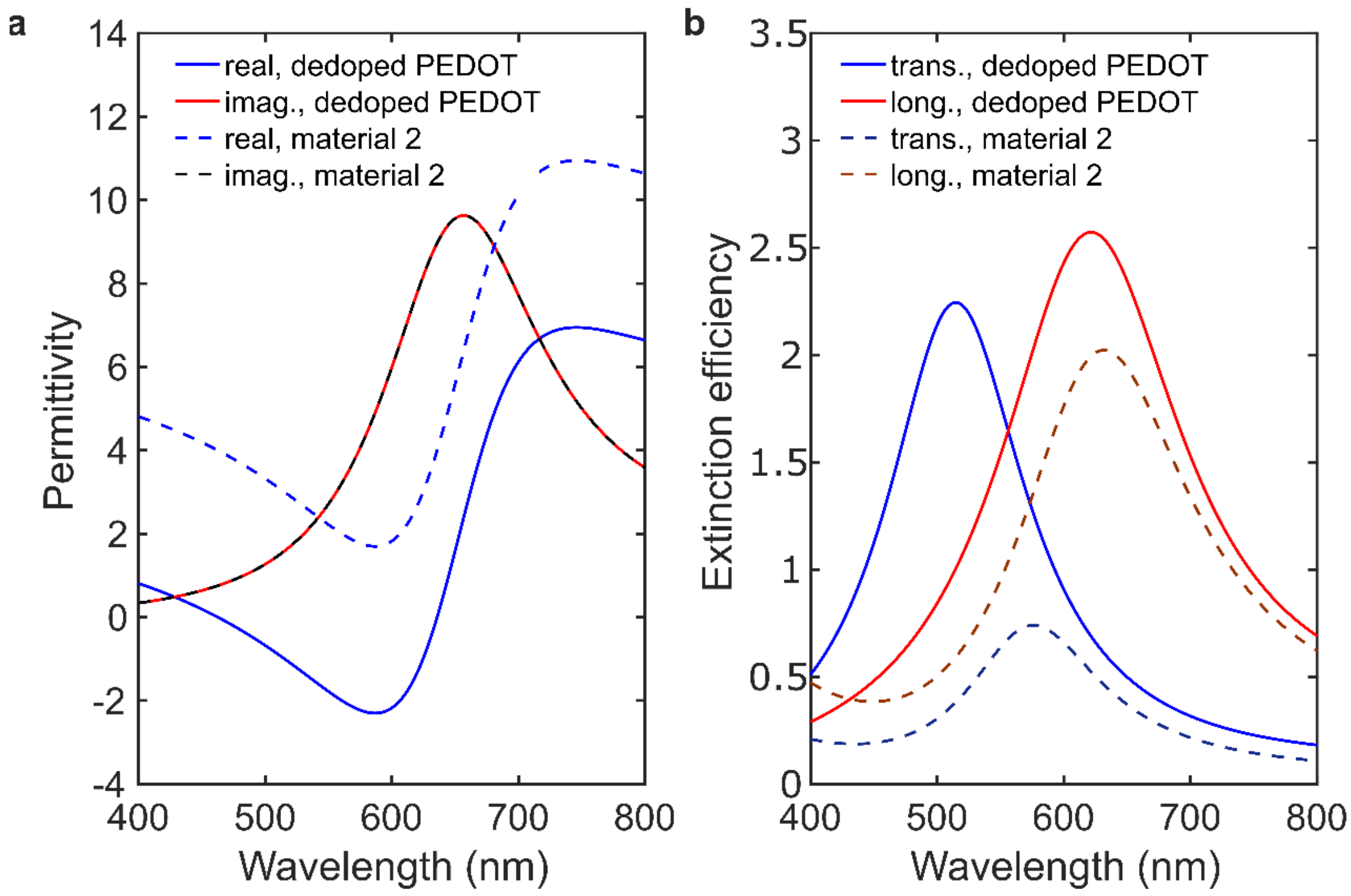

**Extended Data Fig. S8. Influence on background permittivity. (a)** Permittivity of dedoped PEDOT (solid lines) and a material based on PEDOT but background permittivity increased by 4. **(b)** Extinction efficiency for ellipsoids ($l$ = 200 nm, $w$= 50 nm, $h$ = 75 nm) made from these two materials.

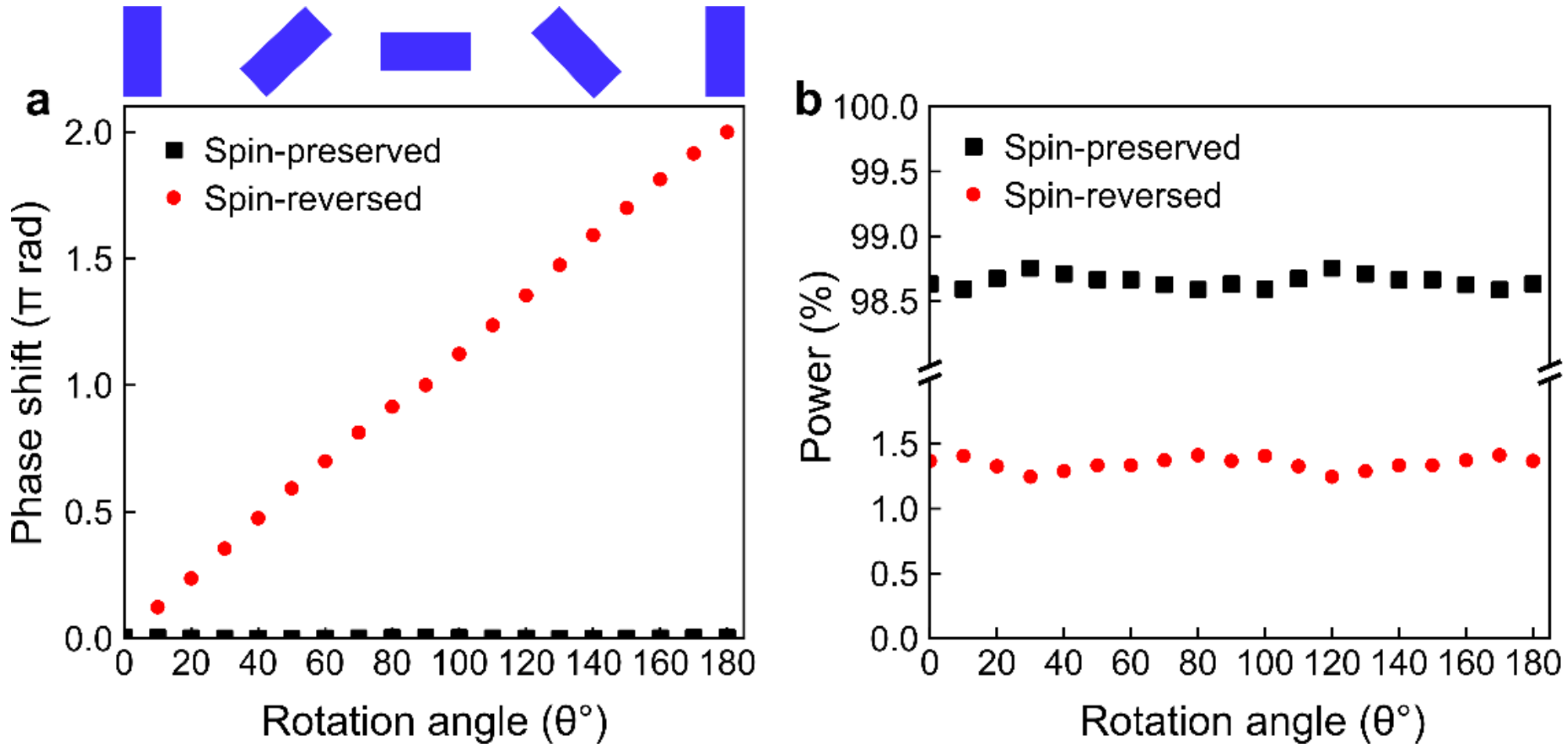

**Extended Data Fig. 9. Phase delay and efficiency for arrays of nanobars of different rotation angles at wavelength of 630 nm. (a)** Phase delay for spin-preserved (black dots) and spin-reversed (red dots) components (with respect to rotation angle 0 set to zero phase shift). **(b)** Transmitted power of the spin-preserved (black dots) and spin-reversed (red dots) components for the same nanobar arrays. The nanobars had dimensions $l$=260 nm, $w$=60 nm, $h$= 70 nm and the period was 450 nm.

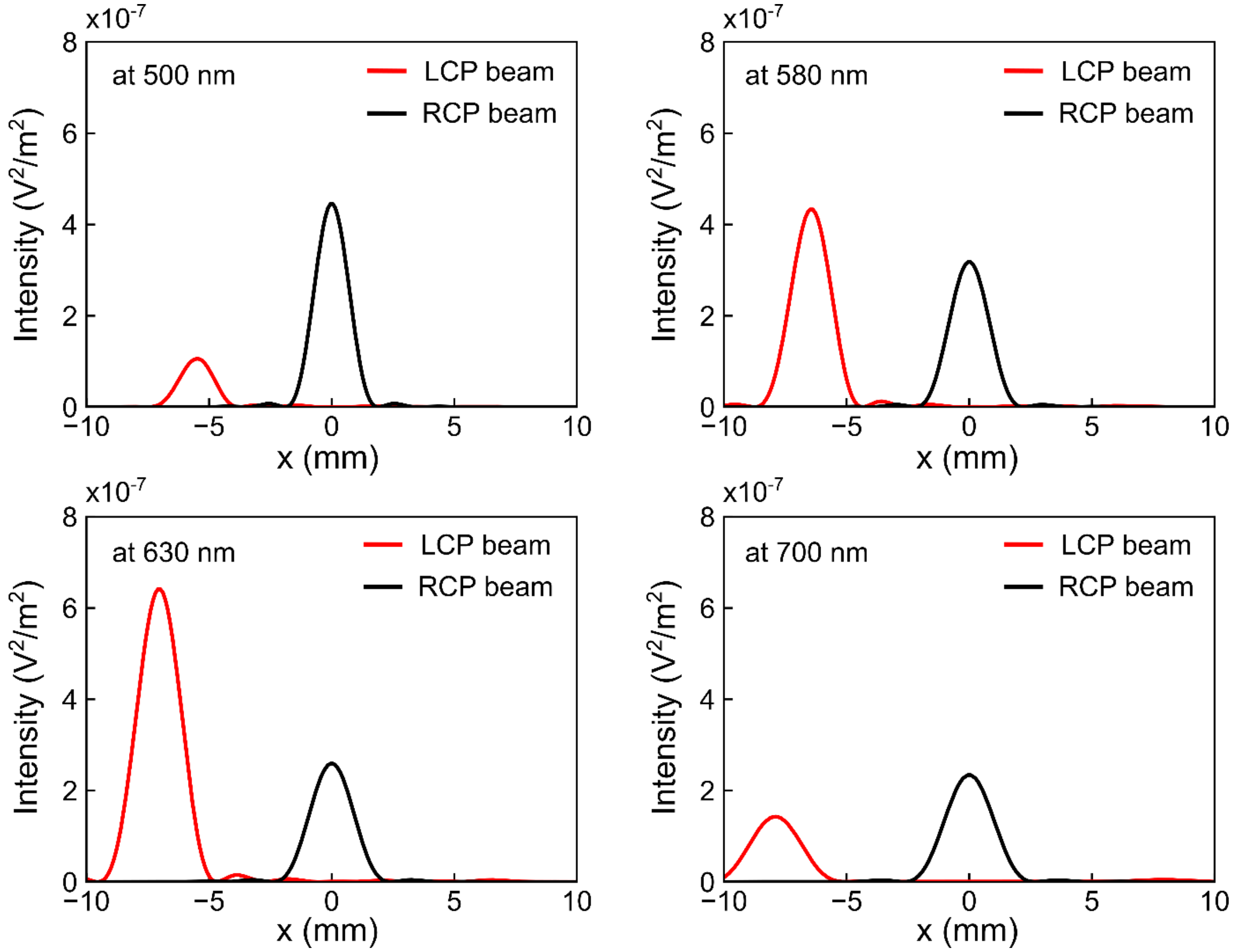

**Extended Data Fig. 10. Simulated far-field beam deflection profile of a 70 nm-thick dedoped PEDOT metasurface at different wavelengths (50 mm projection distance).** The deflection profile of LCP (spin-reversed) is represented by the red line, while the spin-preserved RCP component is shown by the black line. The RCP beam intensity was reduced by a factor of 350 to resemble attenuation introduced by the polarization optics in the experimental setup.

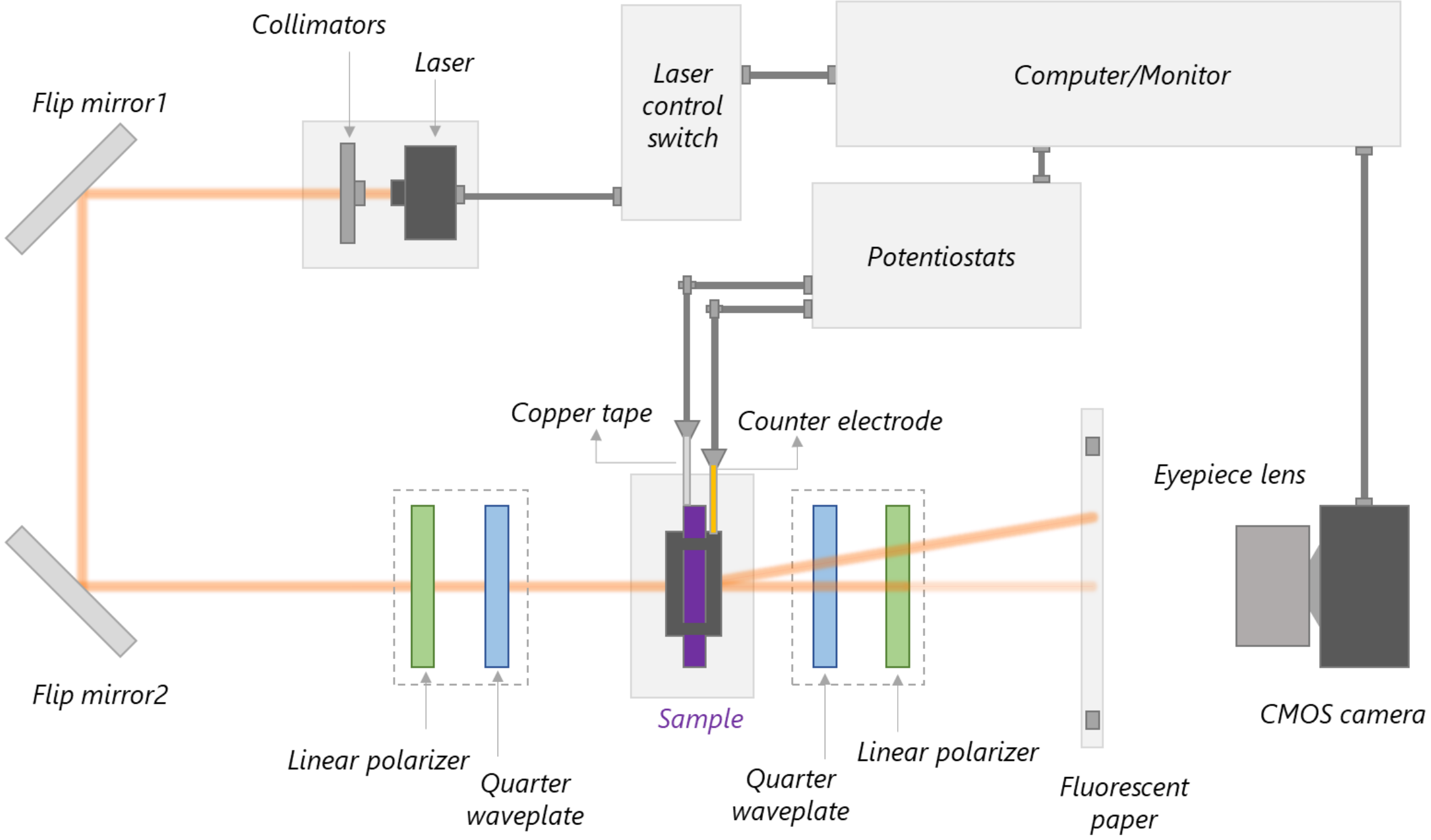

**Extended Data Fig. 11. Schematic illustration of setup used to characterize metasurfaces inducing beam deflection.** A circularly polarized beam is incident on the metasurface, and both the spin-reversed and spin-preserved deflected beams are projected onto a screen placed at a fixed distance of 10 cm.

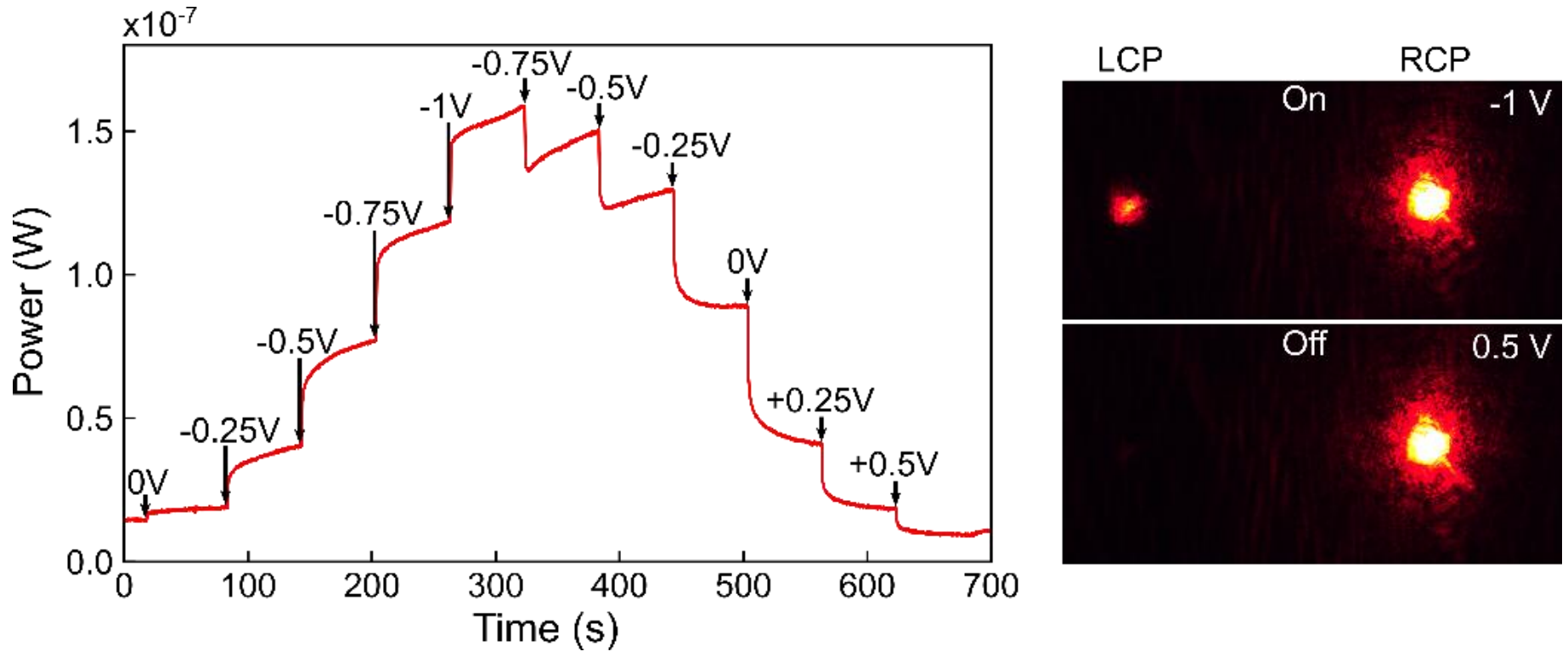

**Extended Data Fig. 12. Measured power of the deflected beam (632 nm) for different doping levels.** Doping levels are controlled electrochemically in steps of 0.25 V, going from 0 V to -1 V to +0.5V. The power of the deflected beam is shown in the left panel. The power of the incident beam was 3.0 mW at 632 nm. The right panel shows a picture of the deflected and incident beam at -1 V (deflected beam on) and +0.5 V (deflected beam off)

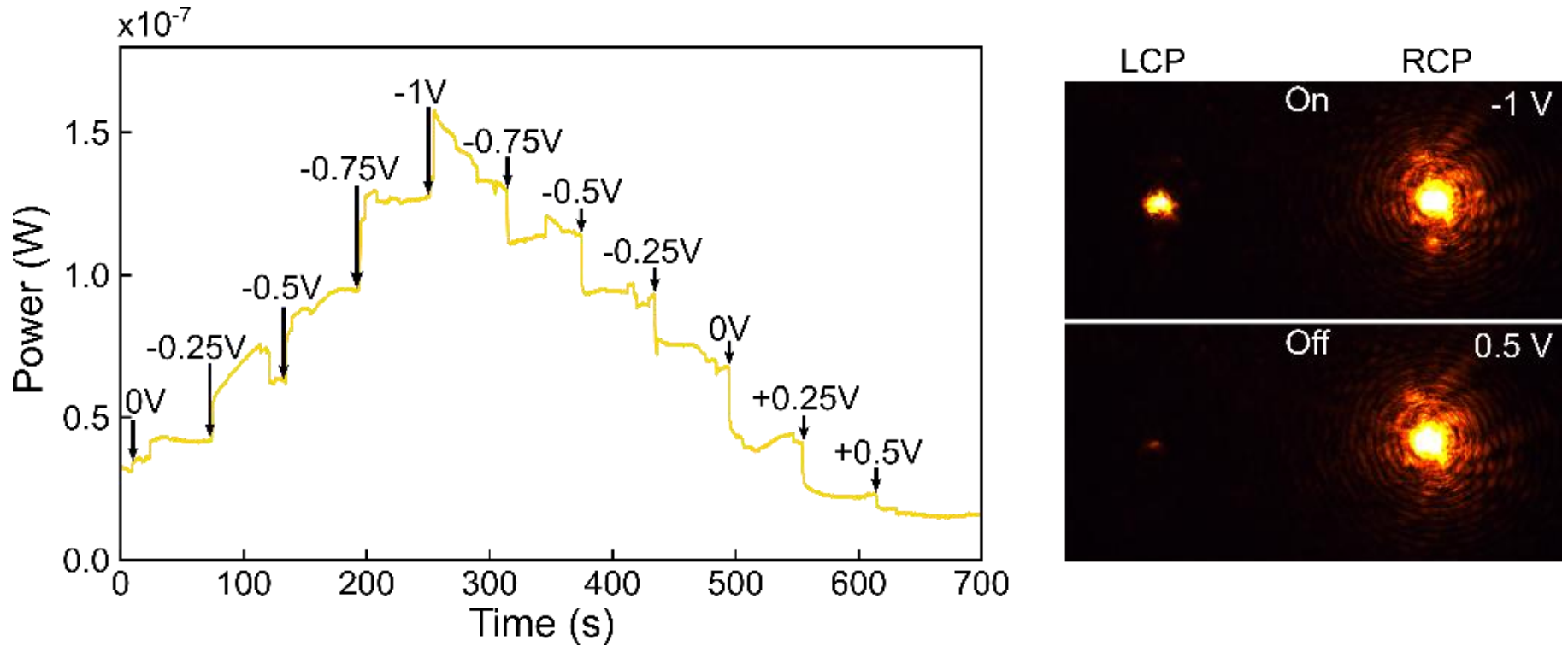

**Extended Data Fig. 13. Measured power of the deflected beam (589 nm) for different doping levels.** Doping levels are controlled electrochemically in steps of 0.25 V, going from 0 V to -1 V to +0.5V. The power of the deflected beam is shown in the left panel. The average power of the incident beam was 9.6 mW at 589 nm, however, it varied between 9.3 to 9.9 mW. The right panel shows a picture of the deflected and incident beam at -1 V (deflected beam on) and +0.5 V (deflected beam off).

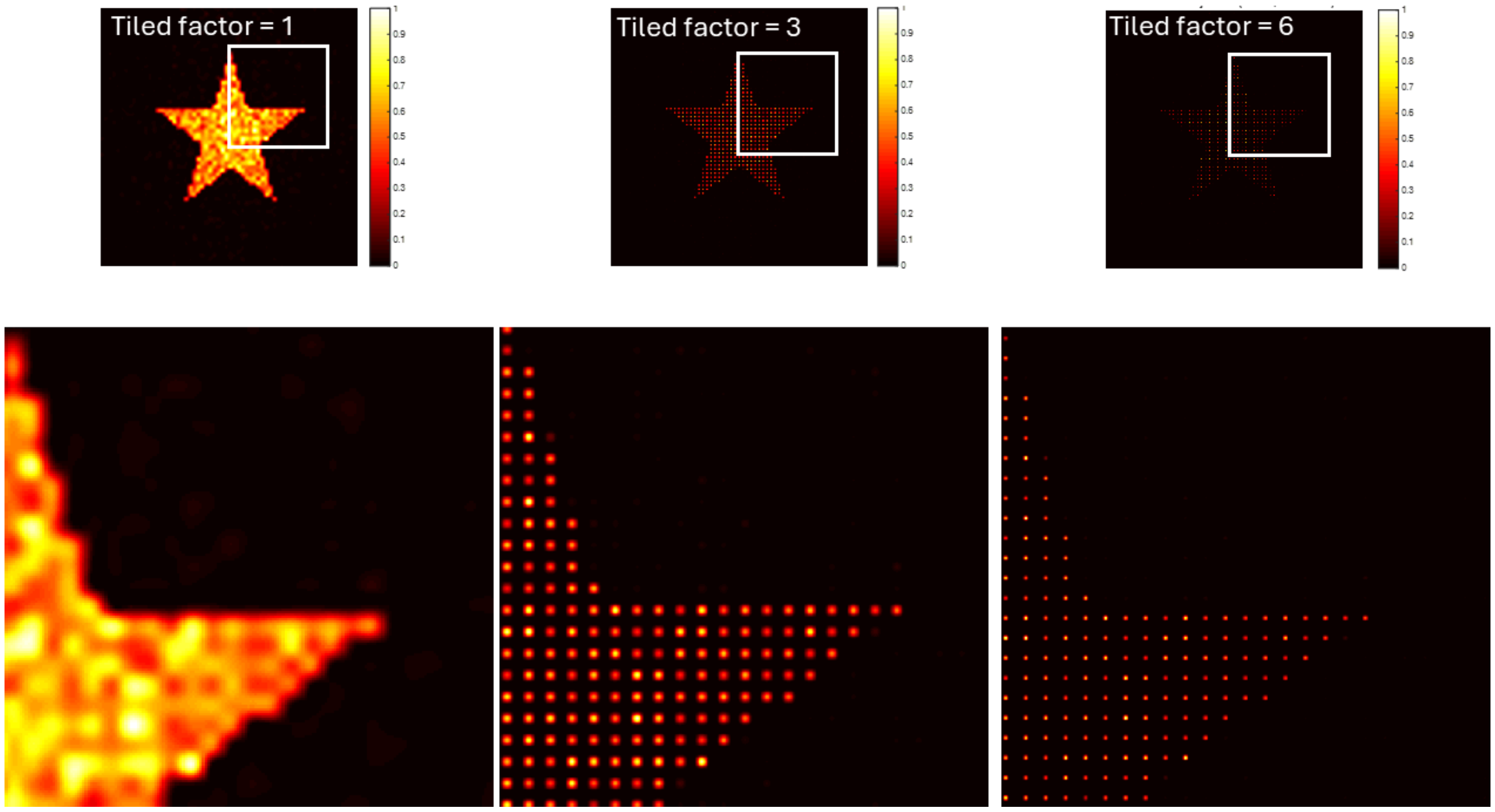

**Extended Data Fig. 14. Calculated holographic images with different tiling factors (left: 1, middle: 3, right: 6) of the 2D phase pattern.** Increasing the tiling factor by repeating the phase map suppresses laser speckle and improves the fidelity and uniformity of the reconstructed holographic image.

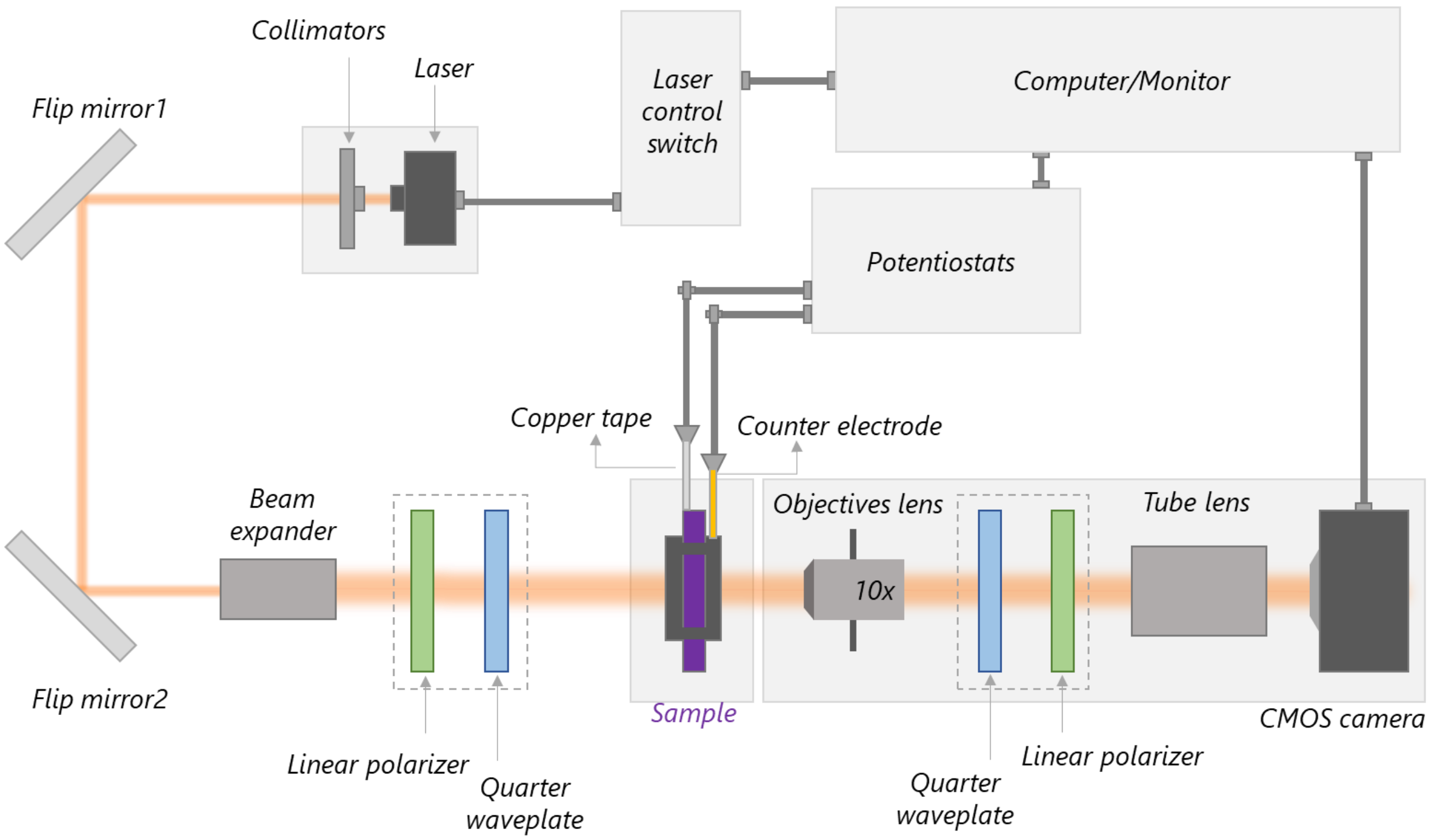

**Extended Data Fig. 15. Schematic illustration of experimental setup for characterization of metaholograms.** A circularly polarized beam is incident on the metasurface, and the reconstructed holographic image appears at a distance of 5 mm from the metasurface. The hologram is collected using 10× objective lens and captured by a camera. The recorded intensity distribution is used to evaluate the reconstruction quality and the metasurface performance under electrochemical switching.